\documentclass[twocolumn]{revtex4}
\usepackage{amsmath,feynmp,graphicx}
\def\bpl{( B \! + \! L )}
\def\bml{( B \! - \! L )}
\DeclareMathOperator{\tr}{tr}
\DeclareMathOperator{\im}{Im}

\begin{document} \setlength{\unitlength}{1in}
    
\title{Leptogenesis with Dirac Neutrinos}

\author{K. Dick$^{1,2}$}
\email{kdick@physik.tu-muenchen.de}
\author{M. Lindner$^{1}$}
\email{lindner@physik.tu-muenchen.de}
\author{M. Ratz$^{1}$}
\email{mratz@physik.tu-muenchen.de}
\author{D. Wright$^{1}$}
\email{ichbin@physik.tu-muenchen.de}
\affiliation{%
$^{1}$Physik Department, 
Technische Universit\"at M\"unchen,
D-85748 Garching, Germany \\
$^{2}$Max Plank Institut f\"ur Physik,
D-80805 M\"unchen, Germany}

\begin{abstract}
\vspace*{0.5cm}
We describe a ``neutrinogenesis'' mechanism whereby, in the presence 
of right-handed neutrinos with sufficiently small pure Dirac masses,
$\bpl$-violating sphaleron processes create the baryon asymmetry of the
Universe, even when $B=L=0$ initially.  It is shown that the resulting 
neutrino mass constraints are easily fulfilled by the neutrino masses
suggested by current experiments. We present a simple toy model which
uses this mechanism to produce the observed baryon asymmetry of the 
Universe.
\end{abstract}

\maketitle

\begin{fmffile}{graphs}

The existence of rapid $\bpl$-violating sphaleron processes 
above the electroweak phase transition \cite{Kuzmin:1985mm} 
tends to wash out any net baryon number which might have 
been produced at higher temperatures and energies.
This washout, it is often argued, makes
higher scale (including GUT-scale) theories of baryogenesis which
begin from $\bml=0$ untenable.
It is important to keep in mind that sphaleron processes do not
directly affect right-handed particles and one might hope that, 
while sphalerons deplete left-handed $\bpl_\mathrm{L}$, right-handed 
$\bpl_\mathrm{R}$ could survive the electroweak epoch and emerge 
afterward as the observed baryon asymmetry of the Universe.
The Yukawa couplings of the SM quarks and leptons to the Higgs, 
however, lead to processes which equilibrate left- and 
right-handed particles so rapidly, that as sphalerons 
destroy left-handed $\bpl$, right-handed $\bpl$ is 
converted to fill the void and is also depleted.
On the other hand, if fermions which carry $B$ or $L$ and have
sufficiently small Yukawa couplings were to exist, then $\bpl$ could
be hidden in right-handed particles until long after the
electroweak phase transition.

In this letter we  point out that neutrinos
with masses as implied by current experiments 
(see e.~g.\ \cite{Zuber:1998xe})
can play exactly such a role. If the neutrinos have pure
Dirac masses, as is for example demanded by $\bml$ conservation, then 
their Yukawa couplings are small enough 
to hide their right-handed lepton number from the sphalerons 
during the entire electroweak epoch.

The mechanism presented here should be called 
\textit{neutrinogenesis}, since the baryon asymmetry of 
the Universe would be a consequence of the smallness of 
neutrino masses. It shares with the well-known 
leptogenesis scenario \cite{Fukugita:1986hr} the exploitation 
of the sphaleron in order to convert a lepton asymmetry 
into a baryon asymmetry, but differs from it in  
important and related aspects. First, instead of 
introducing lepton number violating Majorana masses 
and a see-saw mechanism, this mechanism requires 
only an extended Dirac mass structure in the neutrino 
sector, which implies, however, that the see-saw mass relation can not
be used to "explain" the smallness of neutrino mass.  
The puzzle of the smallness of the Yukawa couplings, which exists
already in the standard model and in models employing the
see-saw mechanism, becomes even more puzzling in the face of the
yet smaller neutrino Yukawa couplings (at least one coupling less
than $10^{-8}$) our scenario demands.
Finally, instead of producing a lepton asymmetry via the decay 
of a heavy Majorana neutrino, this mechanism allows the revival of
old-style out-of-equilibrium decays of massive particles, but
with a twist: instead of producing a baryon asymmetry directly,
the decays should produce a neutrino asymmetry.

\section{Neutrinogenesis}

\begin{center}
\begin{figure}[htb]
\includegraphics*[angle = 0, width = 7cm]{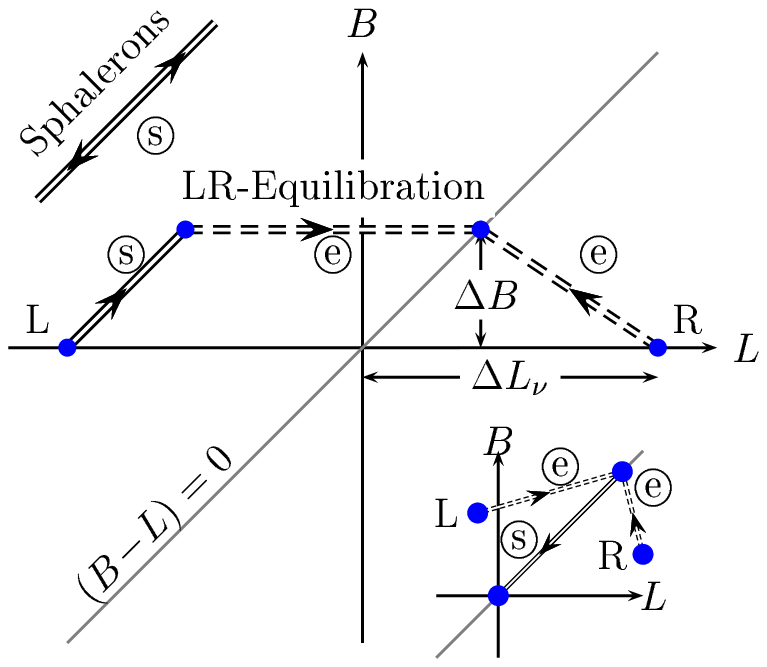}
\caption{\small \label{bild}
Comparison of sphaleronic \textcircled{s} and LR-equilibration 
\textcircled{e} processes affecting $B$ and $L$. For Standard 
Model particles (insert) LR-equilibration occurs completely 
before or during the sphaleron washout. Thus no baryon 
asymmetry can be generated if in total $\bml=0$. For sufficiently 
small Yukawa couplings the LR-equilibration time scale becomes 
longer than the washout period; $\Delta B$ is then generated by 
the sphalerons, even in a theory with $B=L=0$ initially.}
\end{figure}
\end{center}

Let's consider in more detail how the quantity of left- and right- handed 
$B$ and $L$ change in the early Universe, as illustrated in figure 
\ref{bild} for a Universe with $\bml=0$. Initially, some process 
(e.~g.\ at the GUT-scale) produces a total baryon number $B$ and lepton 
number $L$ which are distributed between the left-handed (L) 
and right-handed (R) sectors. Both left-handed and right-handed 
particles are exposed to LR-equilibration processes (marked in 
figure \ref{bild} by \textcircled{e}) which interconvert 
left- and right-handed particles and conserve $B$ and $L$. Sphaleron 
processes (marked in figure \ref{bild} by \textcircled{s}) affect only 
left-handed particles and violate $B$ and $L$ by moving left-handed
$B_\mathrm{L}$ and $L_\mathrm{L}$ along a line of constant $\bml_{\textrm{L}}$.

The interplay of the LR-equilibration and the sphaleron washout
is essentially a comparison of their time scales. For all 
Standard Model particles, the equilibration processes are in 
equilibrium during the epoch in which the 
sphalerons are active.  A baryon asymmetry 
is therefore completely washed out in a theory with $\bml=0$, 
as illustrated in the insert of figure \ref{bild}. The situation is
different with a very weakly coupled right-handed neutrino. Since
LR-conversion \textcircled{e} is not in equilibrium for these particles,
not all $L_\mathrm{R}$ can be depleted; LR-equilibration occurs only after the
sphalerons are ineffective.  Thus only the left-handed components 
are washed out while the right-handed components are preserved,
as shown in the main diagram in figure \ref{bild}.\\

In a more detailed, quantitative picture, we use the
chemical equilibrium of sphaleron and Higgs-mediated reactions
\begin{subequations}
\begin{eqnarray}
                S &\leftrightarrow& 3 q + \ell \\
                \phi &\leftrightarrow& q + \bar{u} \\
                \bar{\phi} &\leftrightarrow& q + \bar{d} \\
                \bar{\phi} &\leftrightarrow& \ell + \bar{e}
\end{eqnarray}
\end{subequations}
and the condition of charge or hypercharge neutrality of the plasma 
to relate the chemical potentials of all the SM and $\nu_{R}$ fields
\cite{Harvey:1990qw}.
One then obtains for $\bml=0$
\begin{equation}
\label{BtoL}
                n_{B} = n_{L} = - \frac{28}{79} n_{\nu_\mathrm{R}}
\end{equation}
in the case of three generations and one Higgs doublet.  This 
confirms the qualitative considerations.\\ 

\section{How (not) to equilibrate $\nu_\mathrm{R}$}

\begin{figure}
\begin{center}
\begin{fmfgraph*}(1,1) 
                \fmfleft{n1,n2} \fmfright{h}
                \fmf{plain,label=$\nu_\mathrm{R}$,label.side=right}{n1,x}
                \fmf{plain,label=$\ell_\mathrm{L}$,label.side=right}{x,n2}
                \fmf{dashes}{x,h}
                \fmfdot{x}
\end{fmfgraph*}
\\[\baselineskip]
\begin{fmfgraph*}(1,1) 
                \fmfleft{n1,n2} \fmfright{l,r}
    \fmf{plain,label=$\nu_\mathrm{R}$}{n1,x}
                \fmf{plain,label=$\ell_\mathrm{L}$}{x,n2}
    \fmf{dashes}{x,y}
                \fmf{plain}{l,y,r}
    \fmfdot{x,y}
\end{fmfgraph*}
\qquad
\begin{fmfgraph*}(1,1) 
    \fmfleft{n1,n2} \fmfright{h,w}
    \fmf{plain,label=$\nu_\mathrm{R}$}{n1,x}
                \fmf{plain,label=$\ell_\mathrm{L}$}{x,n2}
    \fmf{dashes}{x,y,h}
    \fmf{photon}{y,w}
    \fmfdot{x,y}
\end{fmfgraph*}
\end{center}
\caption{\small \label{lr}Processes contributing to the equilibration of $\nu_{R}$.
Dashed lines are Higgs bosons, wavy lines are gauge bosons, and 
unlabeled solid lines are SM fermions.}
\end{figure}
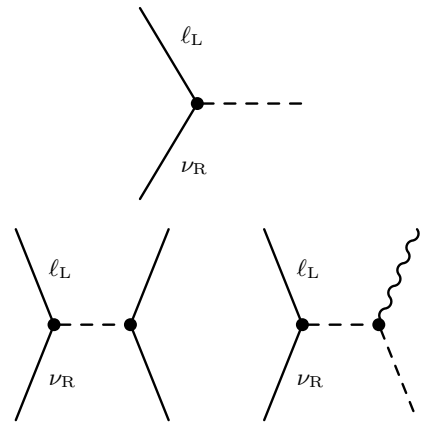

For the success of this scenario, it is necessary that
right-handed neutrinos not be equilibrated quickly above the 
electroweak phase transition.  Schematic Feynman diagrams for the processes
contributing to their equilibration are shown in figure \ref{lr}.  
These processes include Higgs decay and inverse decay, s- and t-channel
scattering off SM fermions, and s- and t-chanel scattering off Higgs 
bosons in combination with the emission or absorption of an electroweak
gauge boson.

The rate of these processes at a temperature $T$ above the electroweak phase 
transition is easily estimated on dimensional grounds to be
\begin{equation}
    \Gamma \sim \lambda^{2} g^{2} T\;,
\end{equation}
where $\lambda$ is the neutrino Yukawa coupling appearing in the 
$\lambda H \ell \bar{\nu}$ term of the Lagrangian and $g$ is a 
gauge or top Yukawa coupling of $\mathcal{O}(1)$.  This rate should be 
compared to the expansion rate of the Universe
\begin{equation}
    H \sim \frac{T^{2}}{M_{\mathrm{Pl}}}\;.
\end{equation}
If $\Gamma > H$ at a temperature $T$ above the electroweak phase 
transition $T_{c}$, left- and right-handed species are equilibrated.
By demanding that this not occur, we obtain the condition
\begin{equation}
    \lambda \lesssim \sqrt{\frac{T_{c}}{M_{\mathrm{Pl}}}}
	\sim 10^{-8}\;,
    \qquad
    m \sim \lambda T_{c} \lesssim 1 \, \mathrm{keV}\;.
\end{equation}
Detailed numerical computation of the corresponding collision terms 
in the Boltzmann equations refines this bound to $\sim 10 \, \textrm{keV}$. 
Although this condition is not fulfilled by electrons, it is easily 
fulfilled by Dirac neutrinos with masses in the range necessary to 
explain Super-Kamiokande, solar neutrino, and LSND data.

\section{Producing $\nu_{R}$: A Toy Model}

\begin{figure}
\begin{center}
\begin{fmfgraph*}(1.2,1) 
    \fmfleft{phi} \fmfright{n,l}
                \fmf{scalar,label=$\Phi$}{phi,x}
                \fmf{fermion,label=$\ell_\mathrm{L}$}{l,y1}
                \fmf{fermion,label=$e_\mathrm{R}$}{y1,x}
                \fmf{fermion,label=$\ell_\mathrm{L}$}{x,y2}
                \fmf{fermion,label=$\nu_\mathrm{R}$}{y2,n}
                \fmffreeze
                \fmf{scalar,label=$\Psi$,label.side=left}{y1,y2}
                \fmfdot{x,y1,y2}
\end{fmfgraph*}
\qquad
\begin{fmfgraph*}(1.5,1) 
    \fmfleft{phi} \fmfright{n,l}
                \fmf{scalar,tension=1.2,label=$\Phi$}{phi,y1}
                \fmf{fermion,left,tension=0.7,label=$\ell_{L}$}{y1,y2}
                \fmf{fermion,left,tension=0.7,label=$e_\mathrm{R}$}{y2,y1}
                \fmf{scalar,tension=1.2,label=$\Psi$}{y2,x}
                \fmf{fermion,label=$\ell_\mathrm{L}$}{l,x}
                \fmf{fermion,label=$\nu_\mathrm{R}$}{x,n}
                \fmfdot{x,y1,y2}
\end{fmfgraph*}
\end{center}
\caption{\small\label{gut}Production of $\nu_{R}$ via decay of GUT scalars.  
The interference of these diagrams with the tree-level decay amplitude 
produces the CP-violation necessary to produce a net neutrino number.}
\end{figure}
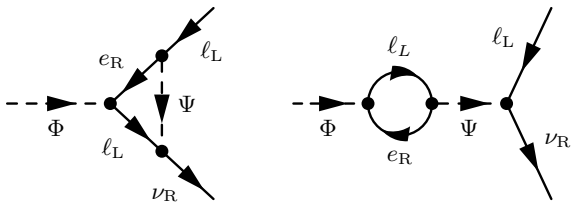

The neutrinogenesis mechanism allows the revival of GUT-scale baryogenesis
by having heavy particles decay into right-handed neutrinos 
instead of directly into particles carrying $B$.
We present here a simple toy model for $B$-generation via this scenario, 
which is loosely based on an old GUT-scale scenario \cite{Fry:1980bc}.  In 
this model, two very heavy SU(2)-doublet scalars (which carry the 
same quantum numbers as the SM Higgs, but will not get vev's)
couple to the SM fields via the Lagrangian
\begin{eqnarray}
 \mathcal{L} &=&
 F(\ell_\mathrm{L}\cdot \Phi)\nu_\mathrm{R}^{c}+
 F'(\ell_\mathrm{L}\cdot \Phi^{c})e_\mathrm{R}^{c}\nonumber\\
&& G(\ell_\mathrm{L}\cdot \Psi)\nu_\mathrm{R}^{c}+
 G'(\ell_\mathrm{L}\cdot \Psi^{c})e_\mathrm{R}^{c}+\mathrm{h.c.}
\end{eqnarray}
They then decay according to
\begin{equation}
        \left. \begin{array}{l} \Phi \\ \Psi \end{array} \right\}
        \rightarrow
        \left\{ \begin{array}{l}
        \bar\ell_\mathrm{L} + \nu_\mathrm{R} \\
        \ell_\mathrm{L} + \bar e_\mathrm{R}
        \end{array}\right.
\end{equation}
Since the elements of $F$, $F'$, $G$ and $G'$ can have relative
phases, interference between the tree-level decay amplitude and 
the one-loop corrections shown in figure \ref{gut} gives rise to a
small CP-violating effect \cite{Nanopoulos:1979gx,Liu:1993tg}:
\begin{eqnarray}\label{DeltaGamma}
\lefteqn{\epsilon_{\Phi} = \frac{
\Gamma(\Phi \rightarrow \bar{\ell} \nu) -
\Gamma(\bar{\Phi} \rightarrow \ell \bar{\nu})}{
\Gamma(\Phi \rightarrow \bar{\ell} \nu) +
\Gamma(\bar{\Phi} \rightarrow \ell \bar{\nu})}} \\
&=& \frac{\im\tr (F^* G F' G'{}^*)}{16\pi \, \tr (F^* F)} \times\nonumber
\\ && \quad
\left[
1 - \frac{M_{\Psi}^2}{M_{\Phi}^2} \ln \left( 1 +
\frac{M_{\Phi}^2}{M_{\Psi}^2} \right) -
\frac{M_{\Phi}^2}{M_{\Phi}^2-M_{\Psi}^2} \right]\;.
\end{eqnarray}
When a bath containing an equal number of $\Phi$ and $\bar{\Phi}$
particles decays, a net
right-handed neutrino number $n_{\nu} = \epsilon n_{\Phi}$ is 
produced, which is not equilibrated as long as the mass condition 
derived in the previous section is fulfilled.  An analogous result holds
for the decays of $\Psi$ particles.

The formalism required to calculate the lepton number stored in 
right-handed neutrinos from such an out-of-equilibrium decay has been 
developed by \cite{Kolb:1980qa,Luty:1992un,kt}, who derive an approximate
expression for the neutrino number to entropy ratio
\begin{equation}
\label{Ynu}
  Y_{\nu} = \frac{n_{\nu}}{s} \sim
 \frac{\epsilon_{\Phi} + \epsilon_{\Psi}}{g_{*}}
\end{equation}
produced in out-of-equilibrium decays.
Here $g_{*} \sim \mathcal{O}(100)$
is the total number of relativistic degrees of freedom in the early
Universe.
To make sure that inverse decays not erase the asymmetry produced in the
decays, we require
\begin{equation}
     K_{\Phi}=\frac{\Gamma(\Phi)}{2H(M_{\Phi})} \sim
     \frac{\lambda^{2}}{g^{1/2}_{*}}
     \frac{M_\mathrm{Pl}}{M_{\Phi}}
     \lesssim 1
\end{equation}
and similarly for the analogously defined $K_{\Psi}$.

In a simplified analysis, one assumes that $M_{\Phi} \sim M_{\Psi} \sim
\mathcal{O}(M)$ and that the largest Yukawa couplings of these scalars are
all $\mathcal{O}(\lambda)$.  In this case
\begin{equation}
\label{simple}
  \epsilon \sim \frac{\lambda^2}{16\pi}\;,
  \qquad\qquad
  \frac{\lambda^2}{g_{*}^{1/2}} \frac{M_\mathrm{Pl}}{M} \lesssim 1\;.
\end{equation}
Combining this estimate (\ref{simple}) with equations (\ref{Ynu}) and
(\ref{BtoL}) with the observed baryon asymmetry
$Y_B = 6$-$8 \cdot 10^{-11}$ \cite{Burles:1999zt}
implies $\lambda \sim 10^{-3}$ and $M \gtrsim 10^{12} \, \mathrm{GeV}$.
This simple estimate shows that a e.~g.\ a weakly coupled GUT could produce the
observed baryon number of the Universe via the neutrinogenesis mechanism.

\begin{figure}
\begin{center}
\includegraphics*[angle = 0, width = 7cm]{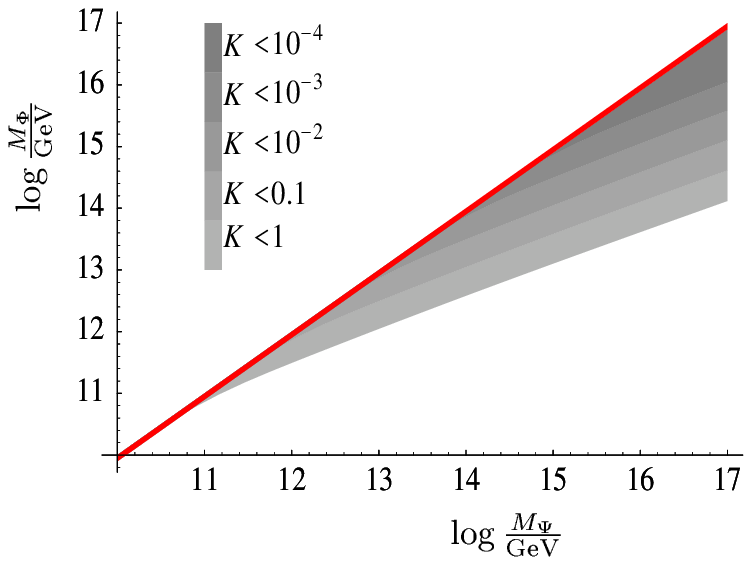}
\end{center}
\caption{\small\label{MassPlot}Allowed masses of the scalars 
of our toy model from the requirements 
for the $K$ values and $n_B/s=8\cdot 10^{-11}$, where we assume 
$M_\Phi <M_\Psi$. The $M_\Phi=M_\Psi$ line has to be excluded since there 
eq. (\ref{DeltaGamma}) cannot be applied.}
\end{figure}

The results of a more detailed quantitative analysis are shown in figure
\ref{MassPlot}, which shows the mass parameters which produce a baryon
asymmetry consistent with observations for various choices
of $K = \max \{K_{\Phi} , K_{\Psi}\}$.

Note that the neutrinogenesis scenario does not require (although it does
allow) decays that violate $B$ or $L$ at the GUT scale; only an asymmetry
between right-handed neutrinos and anti-neutrinos is necessary, since it
alone determines the final baryon asymmetry via equation (\ref{BtoL}).

\section{Discussion and Conclusions}

We presented in this letter a ``neutrinogenesis'' mechanism by which 
$\bpl$-violating sphaleron processes produce a baryon asymmetry instead of 
destroying it, provided that neutrinos have sufficiently small Dirac masses. 
The key observation is that sphalerons couple only to left-handed particles 
while right-handed particles ($SU(2)_\mathrm{L}$ singlets) participate 
in the washout only indirectly via their Dirac Yukawa coupling to 
their respective left-handed partners. The masses of ordinary 
quarks and leptons imply Yukawa couplings for which left--right
equilibration occurs quickly compared to the duration of the sphaleronic 
epoch in the early Universe. However, for Dirac masses below roughly 10~keV,
equilibration takes longer than the washout period. 
Neutrinos with Dirac masses in the experimentally allowed range 
store therefore part of the total lepton number long enough in 
right-handed neutrinos. The sphaleronic washout affects in this 
case only the left-handed neutrinos and $\Delta B$ can be generated 
from the neutrino sector for a theory where initially $\bml=0$ or 
even $B=L=0$. 

In the neutrinogenesis mechanism, baryogenesis is the result of 
an amusing conspiracy of GUT-scale and electroweak-scale effects.
The three Shakarov conditions \cite{Sakharov:1967dj} are realized 
in the following way: CP violating is achieved at some large (e.~g.\ GUT) scale
to produce a neutrino asymmetry. 
Baryon and lepton number are violated only by the electroweak sphalerons 
and both the heavy (e.~g.\ GUT) parents and the light neutrinos are out of 
equilibrium. We presented a toy model which illustrates some details 
and which shows that the right amount of baryon asymmetry can be 
produced for reasonable GUT mass scales. One can easily check that 
constraints coming from Big Bang Nucleosynthesis and from the
matter density in the Universe are not violated. 

Neutrinogenesis should in principle also work in the presence of 
Majorana mass terms for sufficiently small Dirac mass entries. 
One would expect in this case GUT-scale Majorana masses for 
$\nu_\mathrm{R}$ and lepton number would be broken. 
It is also conceivable that neutrinogenesis can be combined in
this way with the known leptogenesis mechanism.
In this case the see-saw mass relation might be used again to 
explain the smallness of neutrino masses.
The most beautiful version of neutrinogenesis is however 
the case of pure Dirac masses and for initial $B=L=0$. 
This could e.~g.\ be realized in suitable GUTs.

A more detailed study of this mechanism and its phenomenology 
will be presented in a longer paper. 

\section*{Acknowledgments}

We thank E. Akhmedov, P. Arnold and M. Yoshimura for helpful discussions.   
This work was supported by the ``Sonderforschungsbereich~375 
f\"ur Astro-Teilchenphysik'' der Deutschen Forschungsgemeinschaft.
One of us (D.W.) gratefully acknowledges the hospitality of the 
Aspen Center for Physics, at which a part of this work was completed.

\end{fmffile}

\bibliographystyle{revtex}
\bibliography{NGL}

\end{document}